\documentclass{iaus}
\usepackage{graphicx}

\title[The NASA {\em EPOXI} Mission] 
{The NASA {\em EPOXI} mission of opportunity to gather ultraprecise photometry of known transiting exoplanets}

\author[Christiansen et al.]   
{Jessie L. Christiansen$^1$,
 David Charbonneau$^1$,
 Michael F.~A'Hearn$^2$,
 Drake Deming$^3$,
 Matthew J.~Holman$^1$,
 Sarah Ballard$^1$,
 David~T.~F.~Weldrake$^1$,
 Richard K.~Barry$^3$,
 Marc J.~Kuchner$^3$,
 Timothy~A.~Livengood$^3$,
 Jeffrey Pedelty$^3$,
 Alfred Schultz$^3$,
 Tilak Hewagama$^2$,
 Jessica M.~Sunshine$^2$,
 Dennis D.~Wellnitz$^2$,
 Don L.~Hampton$^4$,
 Carey~M.~Lisse$^5$,
 Sara Seager$^6$,
 Joseph F.~Veverka$^7$}

\affiliation{$^1$Harvard-Smithsonian Center for Astrophysics, \\ 60 Garden Street, Cambridge, MA 02138, USA\\ email: {\tt jchristi@cfa.harvard.edu} \\[\affilskip]
$^2$University of Maryland, College Park, MD 20742, USA\\[\affilskip]
$^3$Goddard Space Flight Center, Greenbelt, MD 20771, USA\\[\affilskip]
$^4$University of Alaska Fairbanks, Fairbanks, AK 99775, USA\\[\affilskip]
$^5$Johns Hopkins University Applied Physics Laboratory, Laurel, MD 20723, USA\\[\affilskip]
$^6$Massachusetts Institute of Technology, Cambridge, MA 02159, USA\\[\affilskip]
$^7$Cornell University, Space Sciences Dept, Ithaca, NY 14853, USA\\}

\pubyear{2008}
\volume{253}  
\pagerange{119--126}
\jname{Transiting planets}
\editors{A.C. Editor, B.D. Editor \& C.E. Editor, eds.}
\begin{document}

\maketitle

\begin{abstract}
The NASA Discovery mission EPOXI, utilizing the Deep Impact flyby spacecraft, comprises two phases: EPOCh (Extrasolar Planet Observation and Characterization) and DIXI (Deep Impact eXtended Investigation). With EPOCh, we use the 30-cm high resolution visible imager to obtain ultraprecise photometric light curves of known transiting planet systems. We will analyze these data for evidence of additional planets, via transit timing variations or transits; for planetary moons or rings; for detection of secondary eclipses and the constraint of geometric planetary albedos; and for refinement of the system parameters. Over a period of four months, EPOCh observed four known transiting planet systems, with each system observed continuously for several weeks. Here we present an overview of EPOCh, including the spacecraft and science goals, and preliminary photometry results.
\end{abstract}

\firstsection 
\section{Introduction}

The EPOXI mission, led by PI Michael A'Hearn and Deputy PI Drake Deming, is a NASA Discovery Program mission of opportunity, designed to be a low cost, low risk, scientifically focused investigation. The mission comprises of two distinct science projects: Extrasolar Planet Observation and Characterization (EPOCh) and the Deep Impact eXtended Investigation (DIXI). Both investigations make complementary use of the Deep Impact flyby spacecraft, an existing and flight-proven space asset that was previously used to observe comet Tempel 1. The EPOCh project is currently observing known transiting planet systems to obtain ultraprecise photometric light curves and is the focus of this contribution.

The aim of the EPOCh project is to observe a carefully selected set of known, bright, transiting planet systems for several weeks at a time, obtaining light curves with very high precision, phase coverage and cadence. Using these data we can investigate a variety of science goals, including: refining system parameters; searching for additional bodies in the system via transit timing variations in the transits of the known planet (\cite[Agol et al. 2005]{Agol05}; \cite[Holman \& Murray 2005]{Holman05}) or detections of additional transits (\cite[Croll et al. 2007a,b]{Croll2007a}); searching for reflected light at secondary eclipse and constraining the planetary albedo (Rowe et al. and Matthews et al., these proceedings); and searching for the signatures of moons (\cite[Doyle \& Deeg, 2004]{Doyle04}, \cite[Brown et al. 2001]{Brown01}) or rings (\cite[Barnes \& Fortney, 2004]{Barnes04}) associated with the known planet. Another aspect of the EPOCh project was the characterization of Earth as an extrasolar planet (Deming et al. 2007); only the transiting planet observations and results are discussed here.

\section{Observations}

\begin{figure}
\begin{center}
 \includegraphics[width=2.5in]{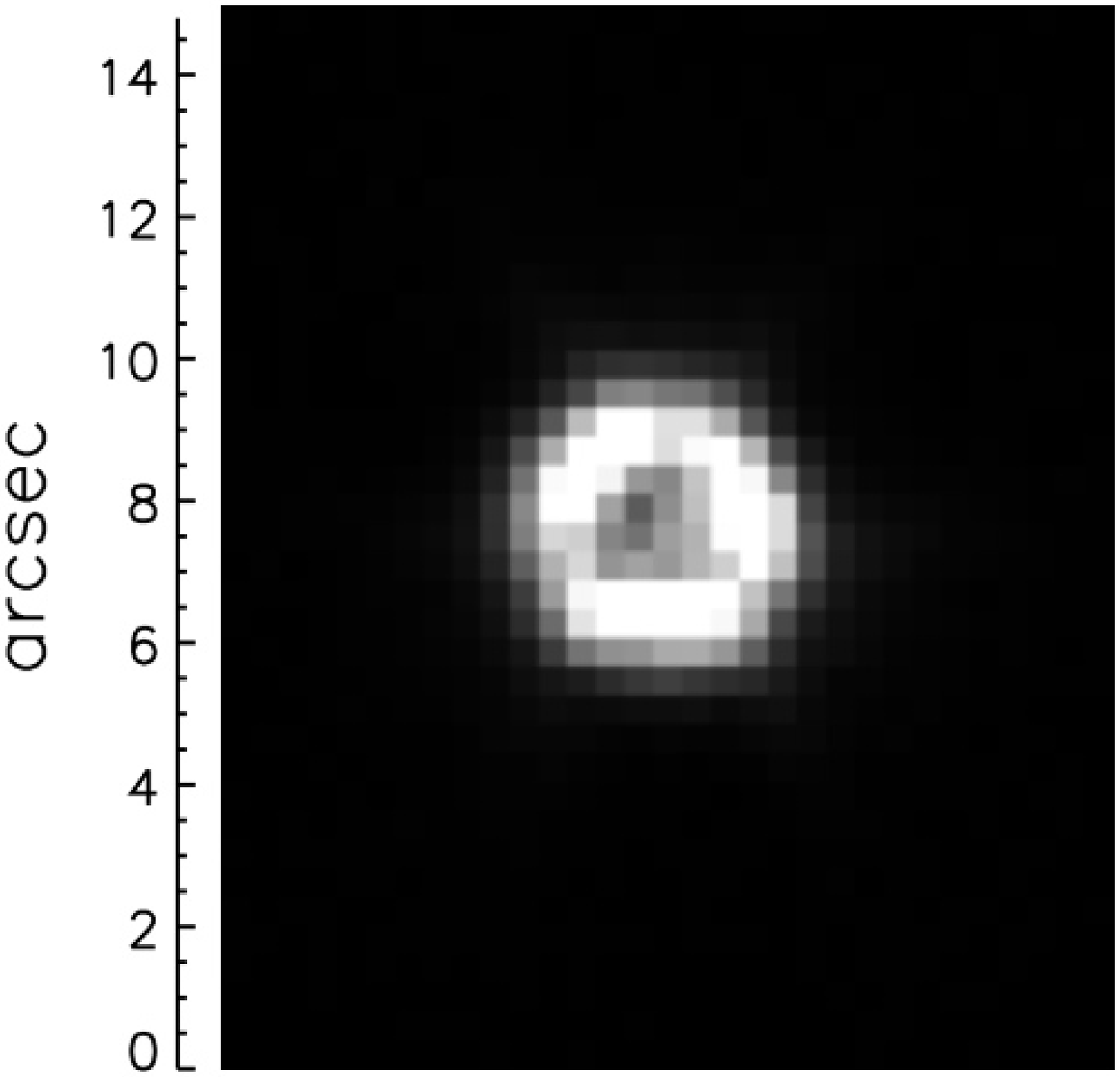}\includegraphics[width=2.9in]{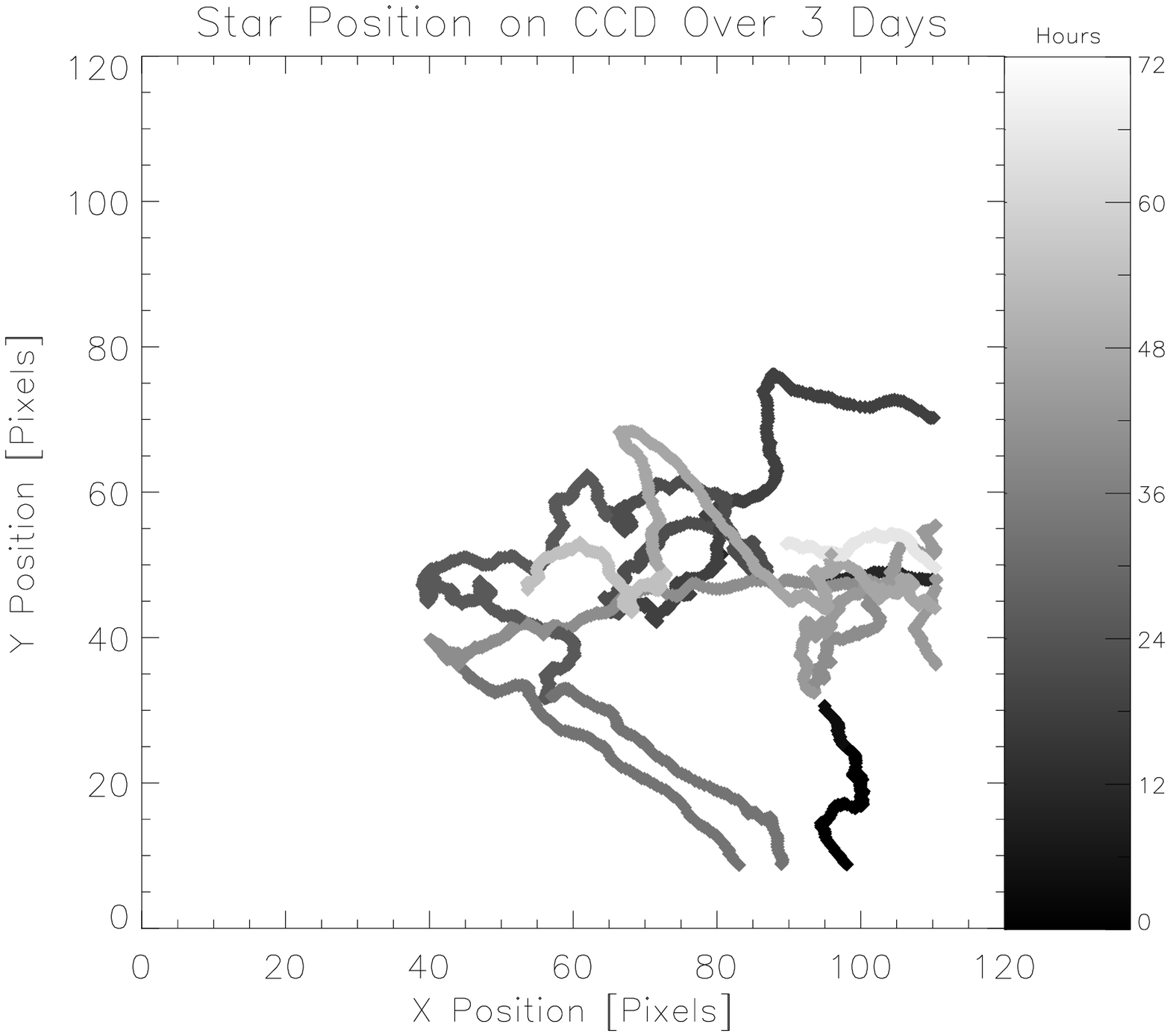} 
 \caption{{\em Left}: The defocused point spread function of the high-resolution imager. {\em Right}: The pointing jitter for three days of HAT-P-4 data. The position of the center of the PSF in the 128$\times$128 sub-array is plotted, where points within 10 pixels of the edge of the chip have been discarded. The color scale indicates the passage of the star over the CCD with time over the 72 hour period.}
   \label{fig:psf}
\end{center}
\end{figure}

We obtained the observations for EPOCh using the high-resolution imager (HRI) on-board the spacecraft. A telescope with a 30-cm aperture illuminates a clear filter with a bandpass of 350--950~nm. The light is imaged in 50 second integrations by a 1k$\times$1k CCD with a pixel scale of 0.4$^{\prime\prime}$ pixel$^{-1}$. Due to the limited amount of on-board memory (230MB), and our desire to maximize the time coverage between data downlinks from the spacecraft, only a 128$\times$128 sub-array of the CCD was initially used for the observations, resulting in a field of view of 51$^{\prime\prime}\times$51$^{\prime\prime}$. The left panel of Figure \ref{fig:psf} shows the highly defocused point spread function (PSF); the advantages for high precision photometry include increased exposure time before saturation and the reduced effect of inter- and intra-pixel sensitivity variations.

The fixed integration time of 50 seconds effectively limited the magnitude range in which our targets could lie to $9 > V > 13$; our final target list and observing schedule are outlined in Table \ref{tab:sched}. These targets were selected to permit a range of scientific investigations.

\begin{figure}
\begin{center}
 \includegraphics[width=5.5in]{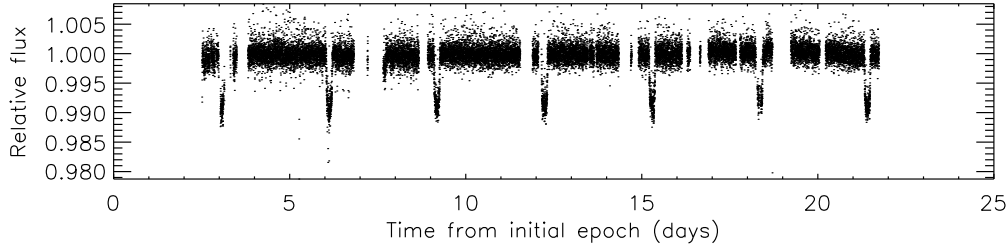} 
 \caption{The full time series of HAT-P-4 photometry, spanning 21 days. The gaps in the time series are due to periods when data are being downloaded from the spacecraft and also when pointing errors resulted in the target moving out of the field of view.}
   \label{fig:hatp4}
\end{center}
\end{figure}

HAT-P-4 (\cite[Kov{\'a}cs et al. 2007]{Kovacs07}) is an interesting system due to its low planetary density, which is unexpected given the high stellar metallicity and the presumed likelihood of the planet having a significant core. We observed seven transits of HAT-P-4, with six having better than 80\% coverage for a duration three times the transit length centered on the time of transit, which is the baseline requirement. The full time series of HAT-P-4 is shown in Figure \ref{fig:hatp4}. 

XO-3 (\cite[Johns-Krull et al. 2008]{Johns-Krull08}) is one of the most massive transiting planets known to date and lies in the relatively unpopulated region between high-mass planets and brown dwarfs. No transits of XO-3 were observed due to the spacecraft traversing perihelion during this time. This increased the on-board temperatures beyond the nominal operating range and the spacecraft entered safe mode. 

TrES-3 (\cite[O'Donovan et al. 2007]{Odonovan07}) has an extremely short period and is an excellent target for reflected light detection, however only 5 transits and 6 secondary eclipses of TrES-3 were able to be observed due to the time lost before the spacecraft was recovered from safe mode.

The spacecraft then experienced a significant decrease in the telecommunications power, and for our next target, XO-2 (\cite[Burke et al. 2007]{Burke07}), which is one of the few known transiting planets to be found in a visual binary star system, we were only able to download partial data for three transits. At this stage observations were suspended until the loss in power could be explained and observations on our penultimate target, TrES-2 (\cite[O'Donovan et al. 2006]{Odonovan06}), were not obtained. TrES-2 is in the \textit{Kepler} field of view (\cite[Borucki et al. 2003]{Borucki03}), and will be observed continuously from space for at least three years by that mission, starting in 2009. Contingent EPOCh observations of TrES-2 (see below) will provide an increased baseline of precise timing measurements with which to probe transit timing variations. 

The system was re-tested when the Earth-spacecraft distance had decreased from the maximum separation, and the power was found to have returned to full downlink capacity. We then resumed observations on our final target in the original schedule, GJ436 (\cite[Gillon et al. 2007]{Gillon07}). This is an exciting prospect for further study due to the non-zero eccentricity of the planet; one explanation is the presence of the third body in the system, pumping up the eccentricity. Seven transits of GJ436 were observed to complete the initial schedule of observations.

Besides the losses due to the spacecraft downtime, there were also significant losses due to pointing errors. Stable pointing is highly desirable for very high precision photometry, and we found that the pointing was insufficiently accurate, in that not only did the initial pointing at the target have a significant offset from the center of the CCD, but the subsequent pointing drifts resulted in the target either leaving the field of view entirely or lingering near the edge, where it is difficult to recover accurate photometry. An example of the pointing wander is shown in the right panel of Figure \ref{fig:psf}, where the position of the center of the PSF on the CCD is plotted for over three days of observations (Rieber \& Sharrow, in prep). Additionally, due to the delay in downloading the data, calculating the required pointing correction and uploading the new coordinates, several days to a week of observations would have been completed in the meantime without correction.

Two strategies were employed to alleviate the pointing problem. The first was the inclusion of a `pre-look', whereby the spacecraft is pointed at a new target, sufficient data obtained to deduce the pointing offset, and the spacecraft then returned to observe the old target while the pointing correction on the new target was calculated. In this way, no observing time was used inefficiently waiting for the upload of the pointing correction. The second alteration was the increase in the dimensions of the sub-array from 128$\times$128 to 256$\times$256 during the predicted transit and eclipse windows, to ensure photometry could be obtained at these critical times. This was put into place for the observations of GJ436 and will be used in the contingent observations discussed below.

\begin{table}
  \begin{center}
  \caption{EPOCh target list and observing schedule.}
  \label{tab:sched}
 {\scriptsize
  \begin{tabular}{l|c|c|l|c}
\hline 
{\bf Target} & {\em V }{\bf Magnitude} & {\bf Period (d)} & {\bf Scheduled}      & {\bf Transits observed}$^1$\\ 
\hline
HAT-P-4      & 11.0                   & 3.06             & 1/22/08--2/12/08      & 6 \\
XO-3         & 9.9                    & 3.19             & 2/13/08--2/19/08      & 0 \\
TrES-3       & 12.4                   & 1.31             & 2/20/08--3/18/08$^2$  & 5 \\
XO-2         & 11.3                   & 2.62             & 3/20/08--4/07/08      & 0$^1$ \\
TrES-2       & 11.4                   & 2.47             & 4/08/08--4/29/08      & 0 \\
GJ436        & 10.7                   & 2.64             & 5/01/08--5/28/08$^2$  & 7 \\ 
\hline
\multicolumn{5}{l}{Contingent observations} \\
HAT-P-4      & 11.0                   & 3.06             & 6/29/08--7/07/08$^3$  & - \\
TrES-2       & 11.4                   & 2.47             & 7/08/08--7/29/08$^3$  & - \\
WASP-3       & 10.6                   & 1.84             & 7/30/08--8/15/08      & - \\
HAT-P-7      & 10.9                   & 2.20             & 8/16/08--8/31/08      & - \\ 
\hline

  \end{tabular}
  }
 \end{center}
\vspace{1mm}
 \scriptsize{
 {\it Notes:}\\
  $^1$ The baseline requirement for a successful transit observation was 80\% coverage. Three partial transits of XO-2 were observed.\\
  $^2$ Earth observations were obtained on 3/19/08, 5/29/08 and 4/06/08.\\
  $^3$ These targets are being returned to in order to fulfill mission baseline criteria.}
\end{table}

The investigation was initially allocated slightly more than four months of observing time, two of which were effectively lost during safe mode and the following telecommunications power loss. In order to recoup the science losses incurred during this period reserve funds were made available to obtain an additional two months of observations beyond the original schedule. These contingent observations are included in Table \ref{tab:sched}. Our plan is to return to HAT-P-4 briefly to complete the mission baseline of five secondary eclipses, and to observe TrES-2, one of the scheduled targets missed in the spacecraft downtime. The final two targets that we will observe are WASP-3 (\cite[Pollacco et al. 2008]{Pollacco08}) and HAT-P-7 (\cite[P{\'a}l et al. 2008]{Pal08}), both very good targets for reflected light detection. The latter is also in the \textit{Kepler} field of view.

\section{Data reduction}

The raw data are converted to calibrated images in a standard fashion (bias- and dark-subtracted and flat-fielded) using the pre-existing Deep Impact data reduction pipeline (\cite[Klaasen et al. 2005]{Klaasen05}). The extraction of the photometry is summarized here and is described in more detail in Ballard et al. (these proceedings). We use a drizzled PSF, oversampled by a factor of 10, to locate the stellar position; bilinear interpolation then results in positions estimated to within one hundredth of a pixel. PSF-fitting to produce the final photometry is complicated by the underlying structure of the CCD, where the central two rows appear to be physically smaller, altering the shape of the PSF when it straddles these rows. We are currently analyzing the data using aperture photometry and scaling each of the rows by a constant factor that is individually determined for each target so as to minimize the rms of the time series.

Initially we found the light curves contained a significant component of red noise, correlated with the star's position on the CCD. The flat fields used in the calibration pipeline were obtained in the lab when the spacecraft was on the ground, and the flat field structure of the CCD may have changed significantly during the flight. An on-board `stimulator' lamp comprised of a green LED mounted near the CCD was used in an attempt to create an updated flat field, however we have found much better results using a decorrelation of the data with position, based on a local 2D spline fit. This has largely removed the red noise component from the light curve; the data shown in Figure \ref{fig:hatp4} are the decorrelated data. The next dominant noise source is the presence of a population of low-energy radiation events, which can be clearly seen in Figure \ref{fig:hatp4} as a preponderance of outliers above the median. The noise in the decorrelated data now decreases with bin size as $\sqrt{N}$, where $N$ is the number of points per bin, as expected for a Gaussian noise profile. 

\section{Preliminary analysis and conclusions}

Our preliminary analysis has thus far focused on the HAT-P-4 dataset; the initial investigations and further work are briefly described below.

\subsection{Refinement of system parameters}

\begin{figure}
\begin{center}
 \includegraphics[angle=90,width=5.7in]{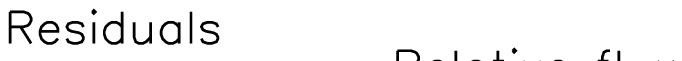} 
 \caption{Phase-folded HAT-P-4 data, overplotted with the best-fit analytic solution from \cite[Mandel \& Agol (2002)]{Mandel02}, with fixed limb-darkening coefficients derived for the instrument bandpass. The data are binned by a factor of 20 and the error bars are calculated using the out-of-transit standard deviation. The residuals to the solution are shown underneath, with a standard deviation of 0.4~mmag.} 
   \label{fig:phase}
\end{center}
\end{figure}

We have used the analytic equations describing transiting light curves of \cite[Mandel \& Agol (2002)]{Mandel02} and the Monte Carlo Markov chain (MCMC) method outlined for radial velocity measurements by \cite[Ford (2005)]{Ford05} and adapted to transit light curve measurements by, e.g., \cite[Holman et al. (2006)]{Holman06} to analyze the light curve of HAT-P-4. We used the four-parameter non-linear limb-darkening law of \cite[Claret (2004)]{Claret04}, with the limb-darkening coefficients derived from a weighted fit across the instrument bandpass, interpolated across the closest theoretical stellar models\footnote{\texttt http://kurucz.cfa.harvard.edu/} (\cite[Kurucz 1994]{Kurucz94}, \cite[Kurucz 2005]{Kurucz05}). The phase-folded binned light curve is shown in Figure \ref{fig:phase} with the best-fit solution over-plotted. Further investigation of the correlation of the errors in the parameters is required, however our results are thus far consistent with those presented in the discovery paper (\cite[Kov{\'a}cs et al. 2007]{Kovacs07}).

\subsection{Reflected light analysis}

\begin{figure}
\begin{center}
 \includegraphics[angle=90,width=5.4in]{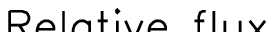} 
 \caption{Searching for the secondary eclipse of HAT-P-4. {\em Upper}: The data around secondary eclipse (set to a phase of 0.0) are shown here, binned by a factor of 80. The solid line shows a model with a secondary eclipse depth inconsistent with the data at a 3$\sigma$ significance, giving an upper limit on the planet-to-star flux ratio of $7.1\times10^{-5}$. {\em Lower}: The change in chi-squared from the best-fit value, with the vertical dashed line delineating the allowed upper limit on the depth.}
   \label{fig:eclipse}
\end{center}
\end{figure}

We have analyzed the data around secondary eclipse in order to search for the signature of reflected light and to put constraints on the geometric planetary albedo. The {\em MOST} satellite has been used very successfully to put tight constraints on this parameter for HD~209458b and HD~189733b (\cite[Rowe et al. 2006, see also the Rowe et al. contribution to these proceedings]{Rowe06}). The upper panel of Figure \ref{fig:eclipse} shows the phased HAT-P-4 data around the predicted times of secondary eclipse (set to a phase of 0.0), binned by a factor of 80. The best-fit eclipse model gives an upper limit on the geometric albedo of 0.346, below that of Jupiter, 0.6. Further work allowing the predicted time of eclipse to vary owing to the range in eccentricity allowed by the radial velocity measurements is still required.

\subsection{Transit timing variations}

\begin{figure}
\begin{center}
 \includegraphics[angle=90,width=5.5in]{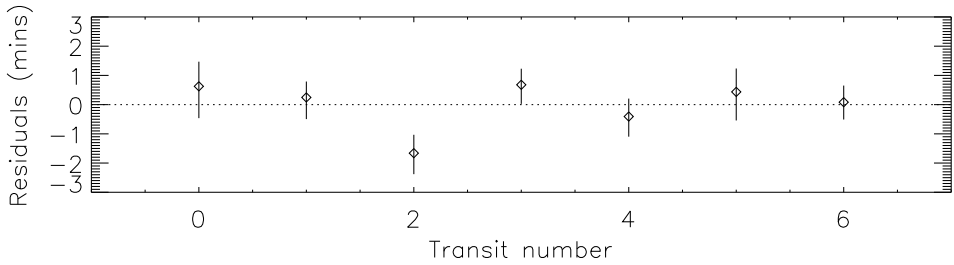} 
 \caption{Residuals of the measured center of each individual transit of HAT-P-4 to the best-fit period.}
   \label{fig:resids}
\end{center}
\end{figure}

The timestamps for the HAT-P-4 data have been corrected to the barycentric Julian date (BJD), however the relative offset between the clock on-board the spacecraft and UT on Earth is yet to be applied. As a result, we can conduct an independent period search and transit timing analysis within the EPOCh dataset, however at this stage we cannot combine these results with those of ground-based observations. We find a period of 3.056370$\pm$0.000094 days, and Figure \ref{fig:resids} shows the residuals of the individual transit times to this period. When the timing offset has been applied, we will analyze the long-term period and transit timing behaviors.

\end{document}